\documentclass[aps,preprint,a4paper,prc,showpacs,showkeys,
nofootinbib]{revtex4-1}
\usepackage{graphicx}
\usepackage{bm} 
\usepackage{amssymb,amsmath,amsfonts}

\begin{document}

\title{On the width of $N\Delta$ and $\Delta\Delta$ states}
\author{J.A. Niskanen}

\affiliation{Helsinki Institute of Physics, PO Box 64, 
FIN-00014 University of Helsinki, Finland }
\email{jouni.niskanen@helsinki.fi}
\date{\today}

\begin{abstract}
It is seen by a coupled-channel calculation that in the 
two-baryon $N\Delta$ or $\Delta\Delta$ system the width of
the state is greatly diminished due to the relative kinetic 
energy of the two baryons, since the internal energy
of the particles, available for pionic decay, is smaller.
A similar state dependent effect arises from the centrifugal
barrier in $L \neq 0$ $N\Delta$ or $\Delta\Delta$ systems.
The double $\Delta$ width can become even smaller than 
the free width of a single $\Delta$. 
This has some bearing to the interpretation of the  $d'(2380)$ resonance recently discovered at COSY.

\end{abstract}


\maketitle

\section{Introduction}
Recently a clear and prominent resonance structure was
observed at the WASA@COSY detector of Forschungszentrum
J\"ulich in double pionic-fusion $pn \rightarrow d\pi^0\pi^0$
\cite{adlar106} and later in isospin associated $pn \rightarrow 
d\pi^+\pi^-$ but not in the isovector channel
$pp \rightarrow d\pi^+\pi^0$ \cite{adlar721}.
Its mass is reported as 2380 MeV, somewhat below two
$\Delta(1232)$ masses, and its width as 70 MeV, and in the
particle zoo it has been nominated as $d'(2380)$. 
The structure is also seen in non-fusion
reactions with isotopically freer four-body final states
$NN\pi\pi$ \cite{adlar88,adlar743}. Thus, along with spin
polarized measurements \cite{adlar112,adlar90}, 
the internal quantum numbers 
$I(J^P) = 0(3^+)$ have been also fixed.

The interpretation of this resonance has been suggested 
as a genuine dibaryon both without \cite{bashka,clement} and 
with explicit quark level calculations 
\cite{huang14,huang15,dong}. Considering that the resonance,
whatever it is, decays mainly through $\Delta\Delta$ it is
understandable that the latter calculations indicate a
dominance of $\Delta\Delta$ in the state wave function 
(about 2/3) and the rest perhaps of more exotic six-quark 
structure. The quota of the six-quark contents would decrease 
the width of the resonance below two times the
 free $\Delta$ width suggested in
Refs. \cite{bashka,clement}. In contrast, a dynamic
three-body calculation \cite{gal1,gal2} can reproduce both
the mass and width without extra explicit quark contents
beyond conventional hadrons, nucleons, $\Delta$'s and pions.
These calculations, however, contained a somewhat fictitious
stable $\Delta'$ to simulate the effect of $\Delta$ in $\pi N$
interaction, which might raise questions about the small width
of the ensuing resonance.

It is the aim of this paper to study in a simple 
phenomenological way the effect of the relative kinetic energy
between the two baryons to see how or if it decreases the
effective decay width of the $N\Delta$ and $\Delta\Delta$ 
two-baryon systems. Obviously this kinetic portion is not 
available for the 
(internal) pionic decay of the $\Delta$'s. Because,
the wave function is necessarily also spatially constrained
(must die asymptotically) the kinetic energy is not arbitrary
and its average is finite. This kinematic suppression of the
width was taken into account long ago in calculations for
$pp \rightarrow d\pi^+$ \cite{improv}, but the width results 
were never explicitly published. Further, also a strong
sensitivity can be expected on the relative orbital angular
momentum of the baryons, which must give rise to quantized
energy levels in closed channels.
Actually a rotational spectrum $\sim
40\,L_{N\Delta}(L_{N\Delta}+1)$ MeV was seen on top of the
$\Delta$ and nucleon mass difference in a coupled
channels $NN-N\Delta$ scattering calculation \cite{dibarmass} 
in good agreement with the isospin one ``dibaryon''
masses given in Ref. \cite{yokosawa}. This would correspond
to a centrifugal barrier height for baryons approximately
at one femtometer distance from each other, roughly the 
distance at which the $N\Delta$ wave function maximizes.

First the system with a single $\Delta$ is treated in
Sec. II to introduce the basic ideas and kinematics before
proceeding to $\Delta\Delta$ of particular interest in the
context of the $d'(2380)$ resonance in Sec. III.

\section{$N\Delta$ states}   \label{single}

In many works (e.g. on pion production reactions 
such as $p + p \rightarrow d + \pi^+$) the effect of the 
$\Delta$ is taken into account
by simply including the $\Delta - N$
 mass difference and width in second order
perturbation calculations into the energy denominator as
$E - \Delta M + i\Gamma /2$. As a trivial consequence,
in gross features this gives the energy dependence 
of the total cross section right, which in this example
around the resonant peak is dominated by a single partial 
wave chain $^1D_2(NN) \rightarrow {^5S_2(N\Delta)} \rightarrow
d\pi^+_{p{\rm -wave}}$, (with also a
significant contribution from
 $^3F_3(NN) \rightarrow {^5P_3(N\Delta)} \rightarrow
d\pi^+_{d{\rm -wave}}$, affecting importantly in differential
and spin observables \cite{ppdpi}). 
In the momentum (or energy) representation this prescription 
is obvious and simple. However, the changes suggested in the
Introduction to the $N\Delta$ kinematics are not necessarily
accounted for. Further,
in different partial waves the centrifugal barrier affects
the magnitude of the $\Delta$ contribution
and can displace the peaking, so that differential 
observables displaying interferences do not come out right
\cite{polarconf}.

\begin{figure}[tb]
\includegraphics[width=\columnwidth]{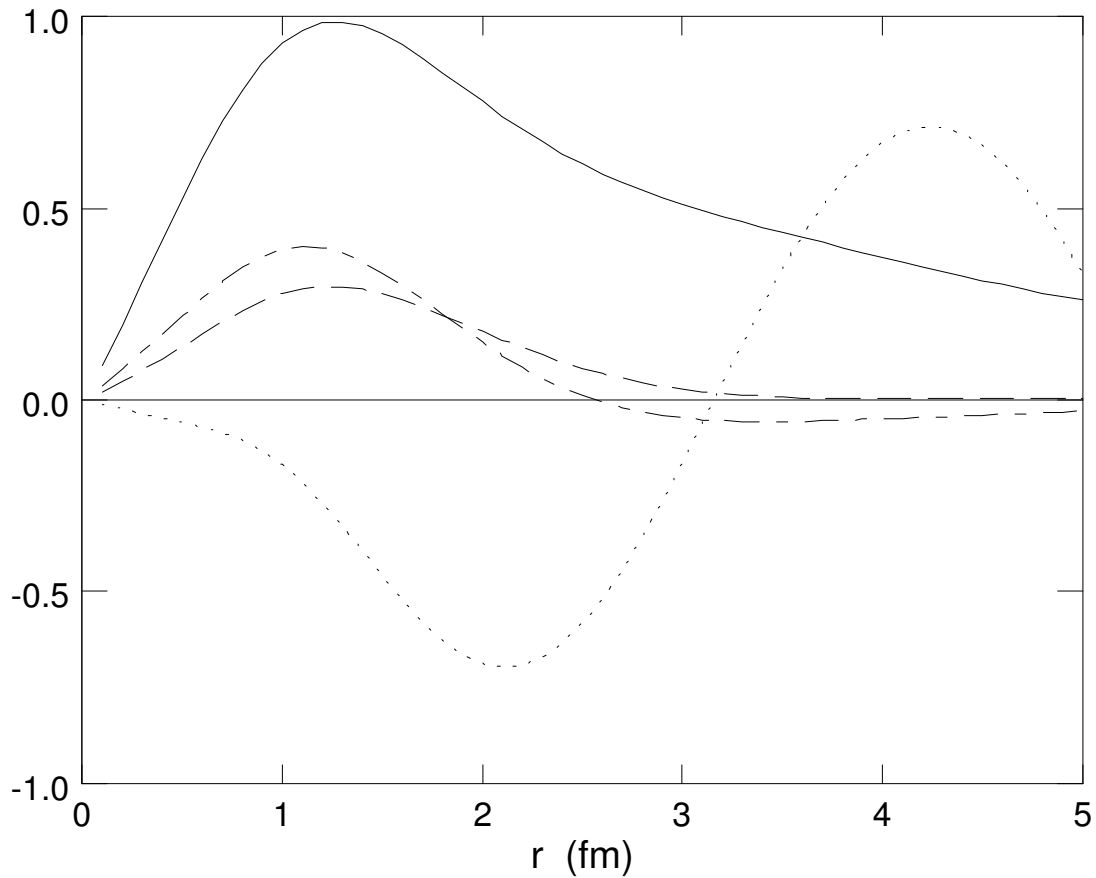}
\caption{The $^5S_2(N\Delta)$ wave function at energies 400,
578 and 765 MeV without the width (dashed, solid and dotted 
curves, respectively). The dash-dot curve has the width included
at 578 MeV. The normalization is associated with the $NN$
wave function asymptotic form $u_{NN}(r) \sim
\sin(kr - \pi + \delta_2)$.
\label{wavef}
}
\end{figure}

As the present calculations are performed in the configuration
space, it is also illustrative to see how the peaking itself
arises with
wave functions obtained from the coupled $NN - N\Delta$
Schr\"odinger equation \cite{ppdpi}. Fig. \ref{wavef} shows 
the most important component $^5S_2(N\Delta)$
of the initial wave
function in the pion production reaction 
$p + p \rightarrow d + \pi^+$. Below the nominal $N\Delta$ mass
this channel is obviously closed and exponentially decreasing
as a function of the distance (dashed curve for 
$E_{\rm lab} = 400$
MeV). At the $N\Delta$ threshold, lacking either positive 
or negative kinetic
energy the wave function becomes essentially a straight curve
outside the potential range $r \geq 2.5$ fm (solid curve at 
578 MeV). Depending on the details of the energy and
the interaction this could, in principle,
be a horizontal constant, maximizing any overlap transition 
integrals (in the case of this reaction with the long-ranged 
deuteron and relatively low energy pion). It may be noted that
already at 600 MeV this line crosses the $r$ axis at
4.6 fm introducing the first oscillation at distances
small enough to cause significant overlap reduction. 
At still higher energies oscillations 
attain shorter wave lengths and begin to cancel the
transition matrix integral (dotted curve).
As a consequence there is a strong peaking of the
production cross section at the $N\Delta$ threshold far 
higher than the data \cite{pl61}.
However, once the $\Delta$ width is included in the equation 
of motion as a constant negative imaginary potential
(as presented in the following discussion), the channel 
becomes again asymptotically closed. As can be seen 
(dash-dot curve at 578 MeV) the wave function becomes 
strongly moderated at short distances and the oscillating 
wave will be exponentially attenuated at large distances 
with a consequent suppression of the transition at and 
above the $N\Delta$ threshold. 
So the natural inputs for the configuration
space equation of motion, the Schr\"odinger equation, lead to
similar resonance like behaviour as can be obtained by explicitly
forcing it by hand in the momentum and energy representation
(see e.g. \cite{riska,brack}).
With the closure of the channels also similar
quantization phenomena appear as for bound states but, however,
smeared with the uncertainties associated with the width. 
As stated previously, the centrifugal barriers (or the
Coulomb force if necessary) can be included with important 
effects on the differential observables with interfering
amplitudes (for $p + p \rightarrow d + \pi^+$ see e.g.
Refs. \cite{ppdpi,improv} vs.  \cite{chai,maxwell1,maxwell2}).

The width is standardly  taken to 
be the free width of the $\Delta$ associated with the
available c.m.s. $NN$ energy. However, as will be seen,
also in this dynamic input quantity  
the effect of the relative $N\Delta$
kinetic energy is significant and dependent on the angular
momentum. This should be subtracted from the internal energy
available for the $\Delta$ decay, so that effectively the
two-baryon width becomes smaller than the "free" width
(which would correspond to zero relative energy of the
baryons).

Further, in these reactions in different partial waves 
also the $N\Delta$ centrifugal barrier directly
diminishes the wave functions. Although this suppression is 
particularly sensitive to $L_{N\Delta}$, even the 
orbital angular momentum of the initial nucleons
may favour transitions into $N\Delta$ in some sense.
Namely, within the interaction range a reduction of the 
centrifugal barrier can compensate the $N\Delta$ mass 
difference in the excitation if $L_{N\Delta} < L_{NN}$,
as seen in Ref. \cite{dibarmass} as an explanation for $T=1$
enhancements ($T=0$ ``dibaryons''). From the above
considerations it is clear that just
a single number cannot account for the effective two-baryon
pole position in different partial waves.

\begin{figure}[tb]
\includegraphics[width=\columnwidth]{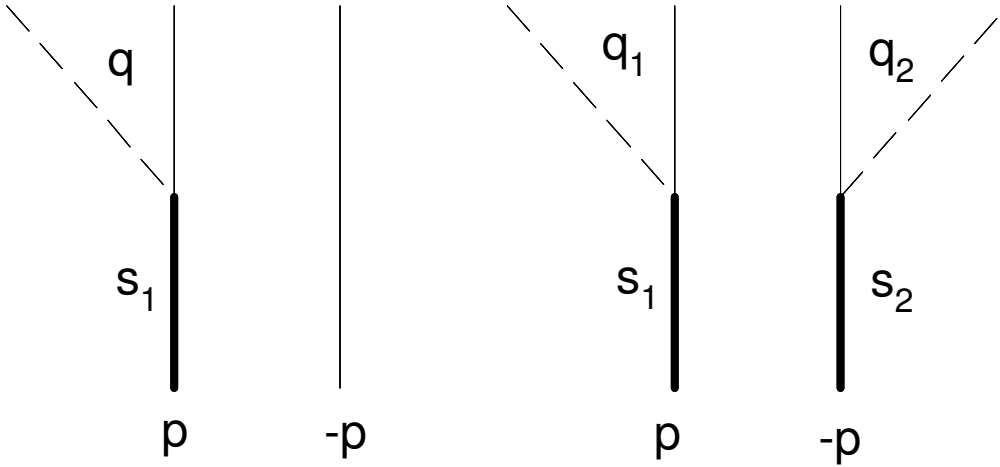}
\caption{Basic kinematics of the $\Delta$ decays. The thick
lines present the $\Delta$'s.
\label{deltas}
}
\end{figure}

Ref. \cite{improv} considered among other things these effects
explicitly by calculating the width into the three-body final
state of Fig. \ref{deltas} as an average over kinematically 
allowed momenta
\begin{equation}
\Gamma_3 = \frac{2}{\pi}\,
\frac{\int_0^{p_{\rm max}} |\Psi_{N\Delta}(p)|^2\,
\Gamma(q)\, p^2\, dp}
{\int_0^\infty |\Psi_{N\Delta}(r)|^2\, r^2\, dr} .
\label{gamma3}
\end{equation}
Here $\Psi_{N\Delta}(p)$ is the Fourier transform of the
appropriate partial wave component of the $N\Delta$ wave
function and $\Gamma(q)$ the free $\Delta \rightarrow N\pi$
width. The maximum relative $N\Delta$ momentum which
still allows the pionic decay is obtained by
\begin{equation}
p_{\rm max}^2 =\, \frac{\lambda(s,(M+\mu)^2,M^2)}{4s}\,
= \,\frac{[s-(M+\mu)^2-M^2]^2 - 4M^2(M+\mu)^2}{4s} 
\end{equation}
from the nucleon and pion masses and the total c.m.s.
energy $\sqrt{s}$. The triangle function $\lambda$ is
introduced in its various forms e.g. in Ref. \cite{byck}.
The physically allowed pion momentum is then constrained by the 
relative baryon momentum through the internal energy of the
$\Delta$
\begin{equation}
s_1 = (\sqrt{s} - \sqrt{M^2+p^2}\, )^2 - p^2
\label{s1}
\end{equation}
to smaller values
\begin{equation}
q^2 = \frac{(s_1 - M^2 -\mu^2)^2 -4\mu^2 M^2}{4s_1}\, .
\end{equation}
Starting with a ``reasonable'' guess for the width(s) the
system is solved iteratively until stable value(s) have been
obtained.

Besides $\Gamma_3$ Ref. \cite{improv} and later 
work with $\pi d$ final states also included the explicit
contribution from this cross section so that the equality
$\Gamma_3 + (\sigma_{d\pi}/\sigma_{\rm tot})\Gamma_{\rm tot}
= \Gamma_{\rm tot}$ was self-consistently satisfied, when
$\Gamma_{\rm tot}$ was used in the coupled-channels calculation
giving $\sigma_{\rm tot}$ as the total inelasticity and
the consequent
baryon wave functions to calculate the $NN \rightarrow
d\pi$ amplitudes. Here the latter term is assumed to be
the two-body ($d\pi$) contribution $\Gamma_2$
to the total width.
Although the present work is not aimed at pion production
{\it per se}, this prescription is nevertheless mainly used 
in this section.
The effect of $\Gamma_2$ is negligible for $NN$ partial 
waves other
than $^1D_2$ and $^3F_3$, where it can contribute about 10--20\%.
It may be noted that, of course, this increases the width 
somewhat and thus acts against the suppression effect claimed here.

Finally, as the free $\Delta$ width input I use a fit to data 
\cite{bransden}
\begin{equation}
\Gamma(q) = \frac{142\, (0.81\, q/\mu)^3}
{1+(0.81\, q/\mu)^2} \; {\rm MeV}
\label{free}
\end{equation}
with the characteristic $p$-wave resonance behaviour 
and a soft form factor.

In addition to the limiting constraints on allowed momenta
in Eq. (\ref{gamma3}), a decisive input necessary is the wave
function of the $N\Delta$ intermediate state, assuming it to
originate from e.g. $NN$ scattering. In this case perturbation
theory with $\Delta$'s is problematic, since there are no 
unperturbed $N\Delta$ wave functions at hand to start with. 
However, the more exact coupled channels approach offers 
probably the best candidates for such wave functions, 
and this method is used here. The coupled
system of Schr\"odinger equations is solved for each incident
nucleon state with the phenomenological Reid potential 
\cite{reid} as the starting point. 
The old age of the interaction does not matter much,
since once the coupling to the excited intermediate 
$N\Delta$ state
is invoked, additional strong attraction is gained, which must,
anyway, be counteracted to avoid double counting. 
This is performed by changing the diagonal $NN$
part so that the total interaction reproduces
the phase shifts \cite{arndt} reasonably well below the
resonance (or $N\Delta$ threshold). 
Ref. \cite{csb} presents such a change to the
most important and sensitive $NN$ states $^1D_2$ and $^3P_1$
(in the original Reid potential)
to be used in this section. An extension of the potential
to the necessary higher partial waves $^3F_3$ and $^3D_3$
is provided by Day in Ref. \cite{day}
\begin{eqnarray}
V(^3F_3) = & 10.463[(1+2/x+2/x^2)e^{-x}/x
  -(8/x+2/x^2)e^{-4x}/x]  \nonumber \\ 
   & - 729.25 e^{-4x}/x  +219.8 e^{-6x}/x  \; , \label{reidf}\\
V_{\rm C}(^3D_3- {^3G_3}) = & -10.463 e^{-x}/x
- 103.4 e^{-2x}/x -419.6 e^{-4x}/x + 9924.3 e^{-6}/x \; , \\
V_{\rm T}(^3D_3- {^3G_3}) = & -10.463[(1+3/x+3/x^2)e^{-x}/x
  -(12/x+3/x^2) e^{-4x}/x ] \nonumber \\
  & + 351.77 e^{-4x}/x  -1673.5 e^{-6x}/x \; , \\
V_{LS}(^3D_3- {^3G_3}) = & 650 e^{-4x}/x -5506 e^{-6x}/x \; ,
\end{eqnarray}
with $x = 0.7 r$. With the $N\Delta$ coupling, as stated above,
these need an additional central repulsion (fitted below the
resonance at 300 and 450 MeV to the energy dependent $pp$ and
$np$ phases of Ref. \cite{arndt} including the most important
$N\Delta$ or $\Delta\Delta$ component)
\begin{equation}
V_{\rm C}({\rm coupled}) = V_{\rm C}(^3F_3) + 2700 e^{-5x}/x \, ,
\;\;
V_{\rm C}({\rm coupled}) = V_{\rm C}(^3D_3- {^3G_3}) + 2800 e^{-7x}/x
\, .  \label{modific}
\end{equation}

The $NN \rightarrow N\Delta$ 
transition potential described in Ref. \cite{improv} has
been fine-tuned to give the height of the $pp \rightarrow
d\pi^+$ peak at the right place $\approx$580 MeV and may be
trusted here, too. This peak is possibly the most 
sensitive probe of the transition potential. 
The potential involves pion and $\rho$-meson
exchanges. The latter may be described by contact terms in 
recent effective field theories, but the main thing in this 
context is to have a transition potential which agrees with
data. The role of the width, in turn, is to act as a constant
imaginary ``potential'' in the $N\Delta$ channels of the
coupled Schr\"odinger equations and produce inelasticity.
It goes without saying that unitarity is not prevailed as 
in general not with optical potentials. At two-baryon level
this is probably closest one can get to reality in the case of
pion production. It is useful to note that besides introducing 
inelasticity the inclusion of the width also acts as effective
repulsion; for moderate inelasticities more imaginary interaction
means less attraction.

\begin{figure}[tb]
\includegraphics[width=\columnwidth]{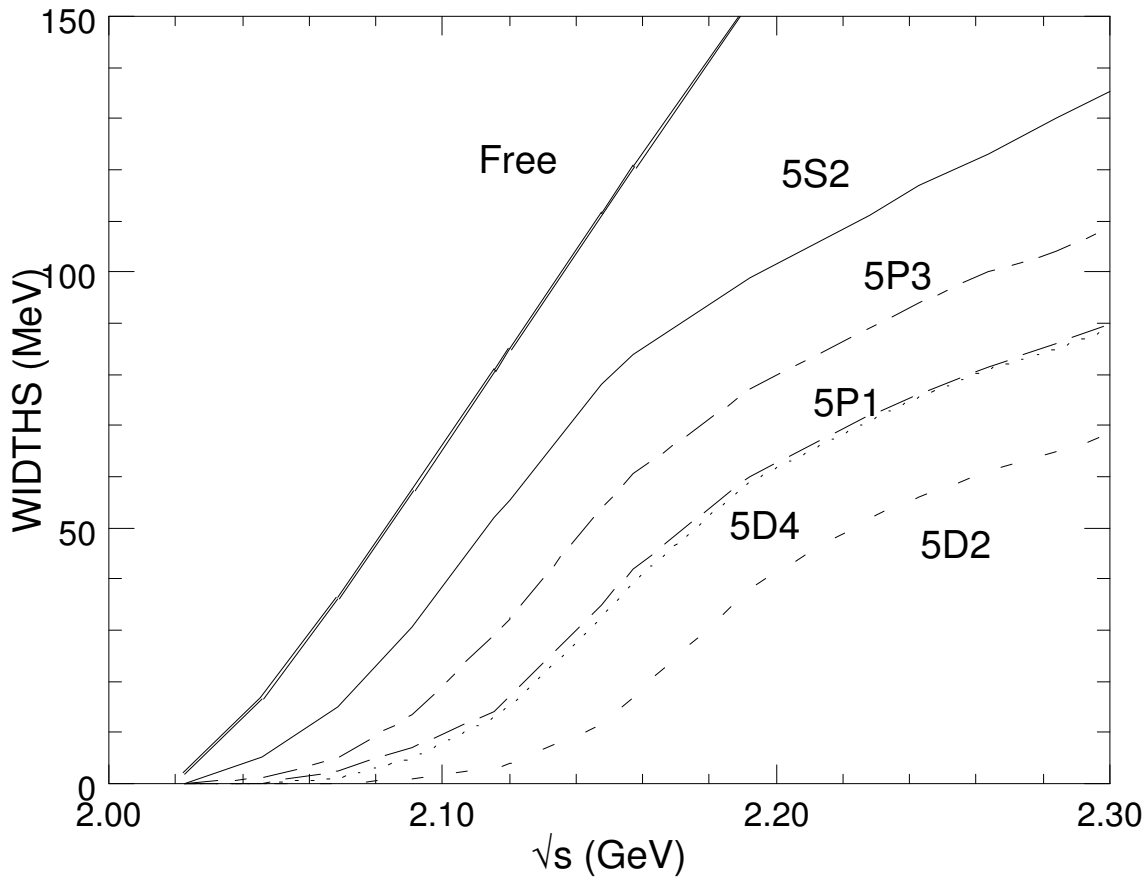}
\caption{The widths of a representative selection of $N\Delta$
states in $NN$ scattering: Curves as described in the text. The
free width is the thick line above the others. 
\label{widths}
}
\end{figure}

Fig. \ref{widths} shows the effective widths of the $N\Delta$
states as functions of the total c.m.s. energy, i.e. the
``dibaryon'' mass for some representative configurations.
For the $NN$ initial states $^1D_2$ and $^3F_3$ the 
criterium for their choice in mainly the importance: The 
orbital
angular momentum decrease in transitions to $^5S_2(N\Delta)$
and $^5P_3(N\Delta)$ favours the formation of $N\Delta$
(solid and dash-dot curves, respectively). A secondary
criterium was to keep the transition potential radially the
same by limiting the discussion to the spin changing tensor 
part and thus minimizing inessential diversions (the 
spin-spin part is excluded in these states). It can be seen
that the $N\Delta$ angular momentum tends to decrease the
width as anticipated earlier. Comparison of the $^5S_2(N\Delta)$
(the highest, thin solid curve) and the $^5D_2(N\Delta)$ 
(the lowest, short dashes) is a 
striking example. It may be noted that they both originate
from $^1D_2(NN)$. The more moderate but clear effect 
of the initial $NN$
angular momentum can be seen between the $^5P_3(N\Delta)$ and
$^5P_1(N\Delta)$ from the $^3F_3(NN)$ and $^3P_1(NN)$
initial states (the dash-dot and dashed curves, respectively).
The straddling of the $^5P_1$
(the dashed curve arising from $^3P_1(NN)$) 
and $^5D_4(NN)$ (the dotted line from $^1G_4(NN)$)
is purely accidental and due to the fact that in the latter
transition the $N\Delta$ orbital angular momentum decreases
from the $NN$, whereas in the former it does not. 
Therefore the $^1G_4(NN)$ is favored as
another $T=1$ ``dibaryon'' \cite{yokosawa,dibarmass}.
For further comparisons also the width of the $^5D_2$-wave
$N\Delta$ is shown by short dashes. First, due to its large
centrifugal barrier this is much smaller than its $S$-wave
sibling. For the reasons already discussed its width is also 
smaller than that of $^5D_4(N\Delta)$,
because the orbital angular momentum does not decrease 
in this transition.
It may be still worth stating that in $^1S_0(NN) \rightarrow
 {^5D_0(N\Delta)}$ the width is numerically negligible, because 
the angular momentum actually increases in the transition.
In the neighbourhood of the $N\Delta$ wave function maximum
the centrifugal barrier is $\approx 200$ MeV, close to the 
$\Delta - N$ mass difference, {\it i.e.} the $N\Delta$
threshold itself.

The thick line presents the free $\Delta$ width (\ref{free})
for static
baryons without any centrifugal barriers. Clearly the 
kinematics of the intermediate baryons have a strong effect
at the nominal mass 2.17 GeV of the $N\Delta$ system and
above even for the $S$-wave $N\Delta$ (thin solid line).
Far above this threshold one might ask about the 
validity of the fit (\ref{free}) to the width, but the softness 
of the form factor should rather underestimate the free 
width than overestimate it.

\begin{figure}[tb]
\includegraphics[width=\columnwidth]{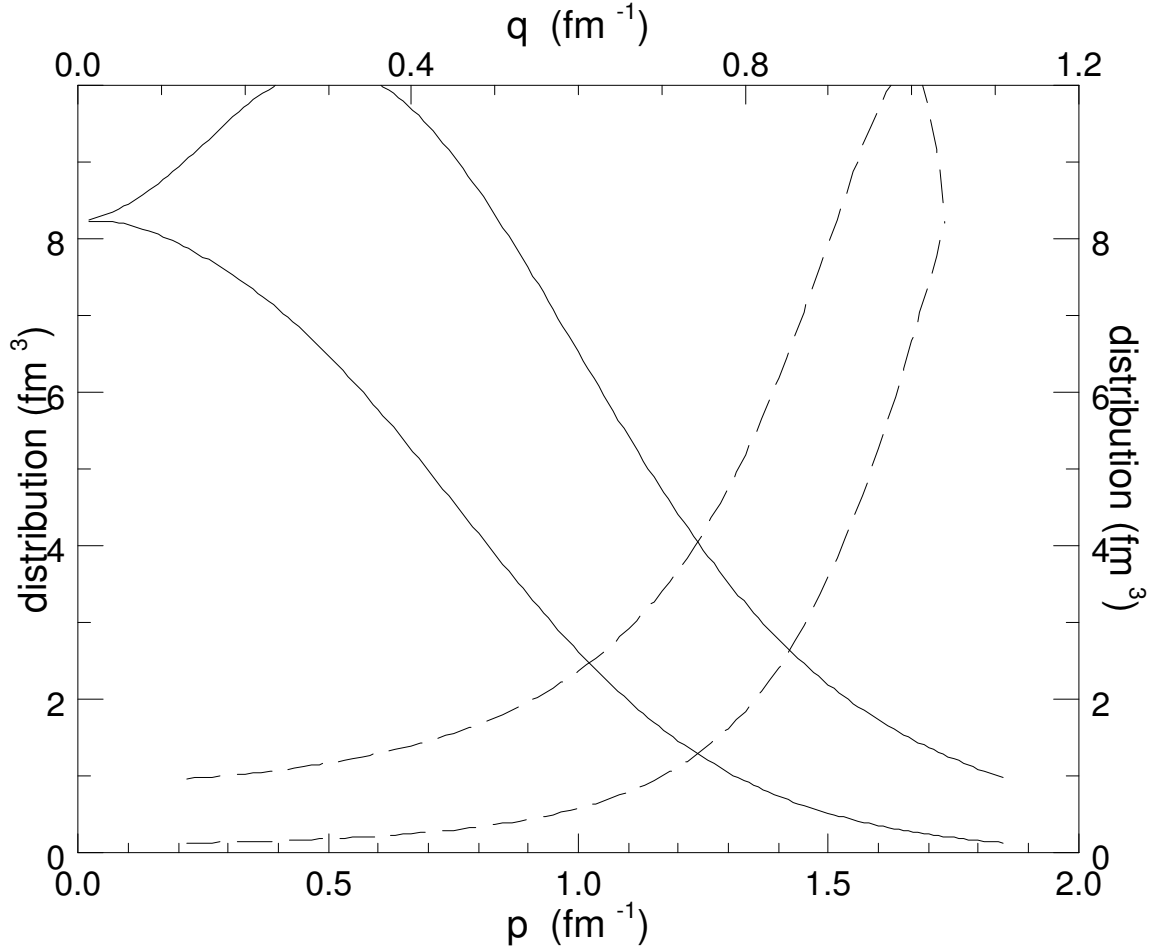}
\caption{Probability distributions of momenta at
$E_{\rm lab} =$ 578 MeV as described in the text; 
the solid curves as functions of the $N\Delta$ momentum $p$ 
and the dashed ones of the pion-nucleon momentum $q$. 
The upper curves are for all main $N\Delta$ contributions 
added together and the lower ones for the dominant
$^5S_2(N\Delta)$ alone.
($p_{\rm max} = 1.87\, {\rm fm^{-1}}$ and
$q_{\rm max} = 1.04\, {\rm fm^{-1}}$). 
\label{distrib}
}
\end{figure}

Sometimes it may be easier or also 
physically more meaningful and beneficial to have the
pion momentum (relative to the recoil nucleon) as the
primary variable. In this case the momentum $p$ is obtained
from
\begin{equation}
p^2 = \frac{(s - s_1(q)-M^2)^2 - 4M^2s_1(q)}{4s}
\end{equation}
and $q_{\rm max}$ (with $s_{\rm 1\, max} = (\sqrt{s} - M)^2$)
from
\begin{equation}
q^2_{\rm max} = \frac{(s - 2M\sqrt{s} -\mu^2)^2 - 4\mu^2 M^2}
 {4(\sqrt{s} - M)^2}   \; .
\end{equation}
In either presentation the probability distribution 
(without the volume
element $\propto p^2$) is given by the absolute square(s) of 
the relevant amplitude Fourier component(s) 
$|\Psi_{N\Delta}(p)|^2$ or $|\Psi_{N\Delta}(p(q))|^2$ 
shown in 
Fig. \ref{distrib} for $E_{\rm lab} = 578$ MeV right at the top 
of the $pp \rightarrow d\pi^+$ cross section \cite{hoftiezer}. 
The partial wave contributions are weighted by the corresponding 
statistical factors $(2J+1)$. The solid curves present the
dependence on $p$ (lower abscissa and left ordinate), whereas 
the dashed ones are for $q$ (upper abscissa, right ordinate). Of
these curves the lower ones include only the  
$^5S_2(N\Delta)$ component coupled to $^1D_2(NN)$), dominant
in $pp \rightarrow d\pi^+$,
whereas the upper ones have all significant smaller components
up to the $^3H_5$ partial wave.
It can be seen that the $S$-wave $N\Delta$ is peaked at small values of $p$, whereas higher angular momentum components 
approach zero there, but are appreciable at higher momenta,
where the kinetic energy of the baryons would be large. 
Of course, the $q$ dependence is opposite to $p$. 
Although the present calculation is not directly aimed at pion
production observables, by {\it e.g.} neglecting the direct
$NN$ contribution, it is conceivable that these contributions
could be seen in pion production into three particles.

\begin{figure}[tb]
\includegraphics[width=\columnwidth]{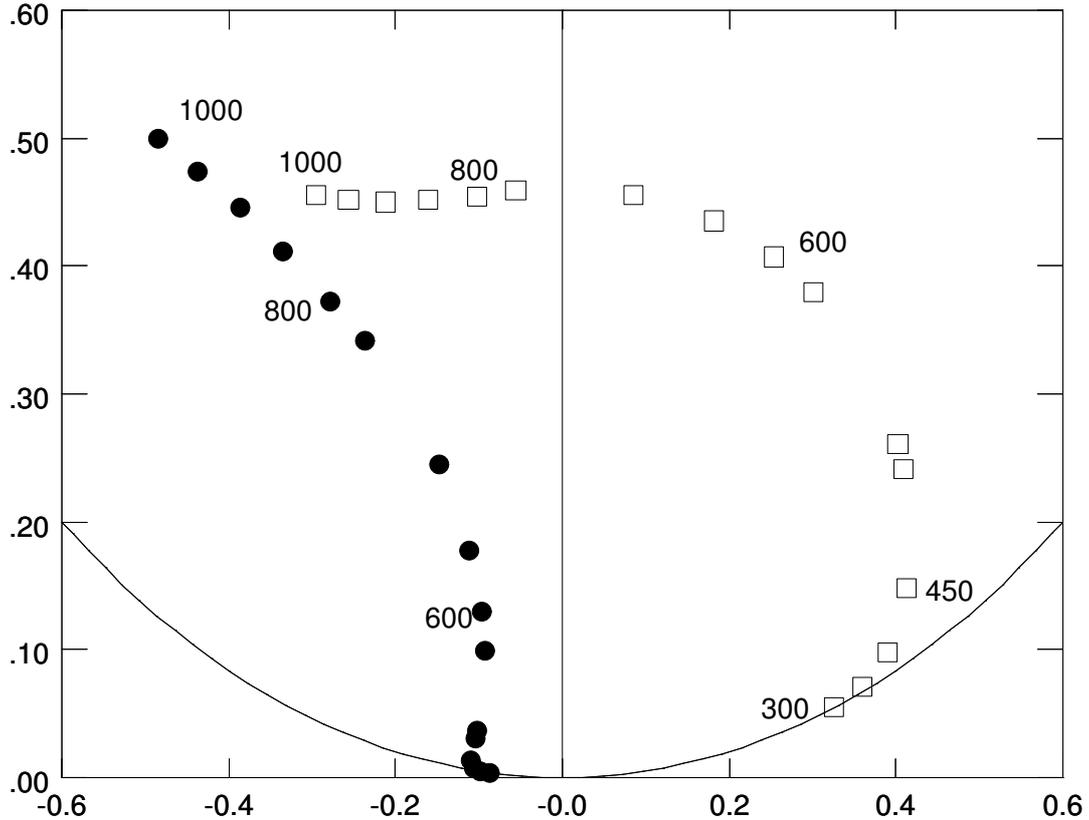}
\caption{Argand diagrams of two $NN$ partial waves. 
Squares $^1D_2$ and full circles $^3F_3$ states. Also
the unitarity circle is shown.
\label{argand}
}
\end{figure}

For a further study of the resonance-like effects of the
$N\Delta$ components Fig. \ref{argand} presents the Argand 
diagrams $2 t = i\, [1-\exp(2i \delta)]$
between $E_{\rm lab}$ 300 and 1000 MeV for 
the $^1D_2$ and $^3F_3$ partial waves, the most
prominent $T=1$ ``dibaryons'', for which the most important
$N\Delta$ configurations were quoted above and in Fig. 
\ref{widths}. Except for the lowest and highest energies 
the mesh is not even spaced but rather follows some 
experimental energies. It can be seen that neither is a full
resonance with the phase passing $\pi /2$ nor do they go around
the center of the unitarity circle. 

\begin{figure}[tb]
\includegraphics[width=\columnwidth]{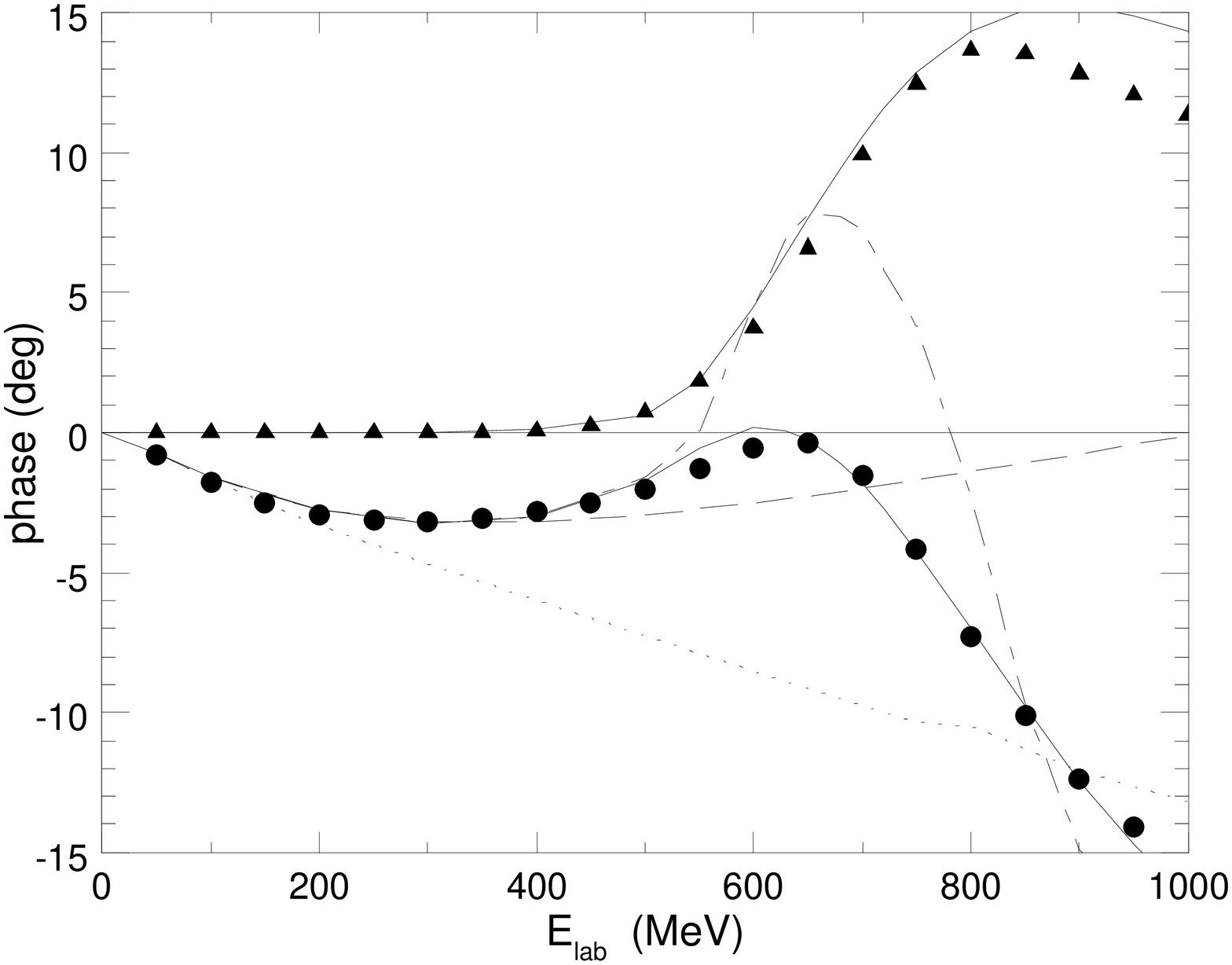}
\caption{The accumulation of the phase shift of the $^3F_3$
state. The Reid potential (\ref{reidf}) result (dashed)
and the modification (\ref{modific}) (dotted). Coupling to the
$^5P_3(N\Delta)$ channel without (dash-dot) and with the
width (solid) as explained in the text. The second solid
curve is the imaginary part of the phase. The data are the
energy dependent fit to $pp$ data from Ref. \cite{arndt}.
\label{f3phases}
}
\end{figure}

So far also the cross section of $NN \rightarrow d\pi$ has been
 accounted for in the widths. To show the effect of a single
$N\Delta$ channel more clearly I constrain in the discussion
to $NN(^3F_3) \rightarrow N\Delta(^5P_3) \rightarrow NN(^3F_3)$
neglecting also the $F$-wave $N\Delta$'s. This results in about 
5\% decrease in the width from that shown in Fig. (\ref{widths}).
Altogether the neglect of the $d\pi$ and the $F$-wave $N\Delta$'s
is a loss of less than one degree of attraction at intermediate
energies of interest here. The latter neglect has practically no
effect on the $P$-wave width.

In Fig. (\ref{f3phases})
the accumulation in the phase shift $\delta(^3F_3)$ arising
from the Reid potential (\ref{reidf}) and the coupling to 
only $^3P_3(N\Delta)$ is presented. First the potential itself
gives a flat and relatively featureless result, which however
agrees excellently with the analysis \cite{arndt} up to the
pion production threshold (dashed curve). The modification 
(\ref{modific}) is too unrealistically repulsive (dotted)
but due its very short range does not change 
the low energy agreement much.
However, the coupling to the $N\Delta$ state returns the
attraction but without the width leads to a very narrow and too 
high peak at $\approx 660$ MeV (dash-dot) slightly above the
$N\Delta$ threshold and well in accordance with the prescription 
\cite{dibarmass} quoted in the Introduction. 
Finally the inclusion
of the width smooths the peak and
gives the solid curve in good agreement with data
up to one GeV. Actually the deviation from the data is less than 
or of the same magnitude as the difference between the $pp$ and
$np$ analyses. Also the imaginary part of the phase shift is
in reasonable agreement with the data extracted from the 
$K$-matrix of Ref. \cite{arndt} (triangles).
It is also interesting and illuminating to note
that about 100 MeV above the nominal $N\Delta$ threshold (center
of mass) the coupling effect turns repulsive (the solid curve
gets below the dotted) showing typical threshold cusp (or
resonance) behavior. However, the smooth background potential
repulsion keeps the phase negative and thus the corresponding 
Argand diagram remains on the left side of the imaginary axis
in Fig. \ref{argand}. In the partial wave cross 
section the phase shift maximum here should then
show rather as a minimum than as the ``standard''
maximum. Of course, this minimum in $pp$ scattering has not
much to do for the $NN \rightarrow d\pi$ reaction, where
$^3F_3$ is the second state in importance besides $^1D_2$
above the threshold region, but this importance is based on
the overlap of the $N\Delta$ configurations with the final
$d\pi$ states -- mainly $^5P_3(N\Delta)$ and $d$-wave pions.

\section{$\Delta\Delta$ states}

Conceptually the width of a single $\Delta$ even in presence 
of another nucleon is quite clear. For a pair of $\Delta$'s
the situation is slightly more complex. Some works in the
context of the $d'(2380)$ have considered twice the single
$\Delta$ width as relevant \cite{bashka,clement}. 
However, it is difficult to see why the lifetime of two 
$\Delta$'s should be only half of the lifetime of a single 
$\Delta$. Rather, if one considers as the lifetime
the time that is required for {\it both} to decay, by
conditional probabilities the lifetime in this sense should
be longer and the width smaller. After all, the experimental 
results are for the two decays with two pions.

In the latter sense one might start from the probability for
the two unstable particles
to decay, one after the other (in both orders)
$\sim \exp(-\Gamma_1 t_1) \exp(-\Gamma_2 t_2)$ and perform
the double time integral. This would give a time dependence 
for the transition $\sim \exp(-\Gamma t)[\exp(-\delta t) +
\exp(+\delta t)- \exp(-\Gamma t)]$ with $\Gamma$ the average
$(\Gamma_1 + \Gamma_2)/2$
of the widths and $\delta = (\Gamma_1 - \Gamma_2)/2$. This
dependence is dominated by the average, which for the two
$\Delta$'s would be simply the single normal width. From the 
kinematic results of Sec. \ref{single} it might be possible 
that even this is further decreased. However, with the energy
scale of the double $\Delta$ it is also possible that the 
``free''
width $\Gamma(q)$ could get very large values for large 
momenta and the ensuing integrals would yield larger widths
instead. In the absence of firm intuitive arguments  
an explicit estimation is required.

Now the two-$\Delta$ width is calculated as the double
integral
\begin{equation}
\Gamma_4 = \frac{2}{\pi}\,  \frac{\int |\Psi_{\Delta\Delta}(p)|^2
[\Gamma(q_1) + \Gamma(q_2)]/2 \, p^2dp\, dq_1}
{q_{\rm max}\, \int_0^\infty |\Psi_{\Delta\Delta}(r)|^2\, r^2dr}
 \; .
\end{equation}
Here the maximum limit of the free variable $p$ is obviously
from the kinematics of Fig. \ref{deltas}
$p_{\rm max} = \sqrt{s/4 - (M+\mu)^2}$ and the upper limit of
the pion momentum as a function of $p$ is obtained from the
maximum internal energy of particle one
\begin{equation}
s_{\rm 1max} = (\sqrt{s} - \sqrt{(M+\mu)^2+p^2}\;)^2 - p^2
\end{equation}
as
\begin{equation}
q_{\rm 1max}^2 =  \frac{(s_{\rm 1max} - M^2 - \mu^2)^2
- 4\mu^2M^2}{4\, s_{\rm 1max}} \; .
\end{equation}
In the pion integration the second dependent momentum $q_2$
in turn is obtained from
\begin{equation}
q_2^2 = \frac{(s_2 -M^2 -\mu^2)^2 - 4\mu^2M^2}{4\, s_2}
\end{equation}
with $s_2 = (\sqrt{s} - \sqrt{s_1+p^2}\;)^2 - p^2$ and
$s_1 = (M^2+q_1^2) + (\mu^2+q_1^2)$.

\begin{figure}[tb]
\includegraphics[width=\columnwidth]{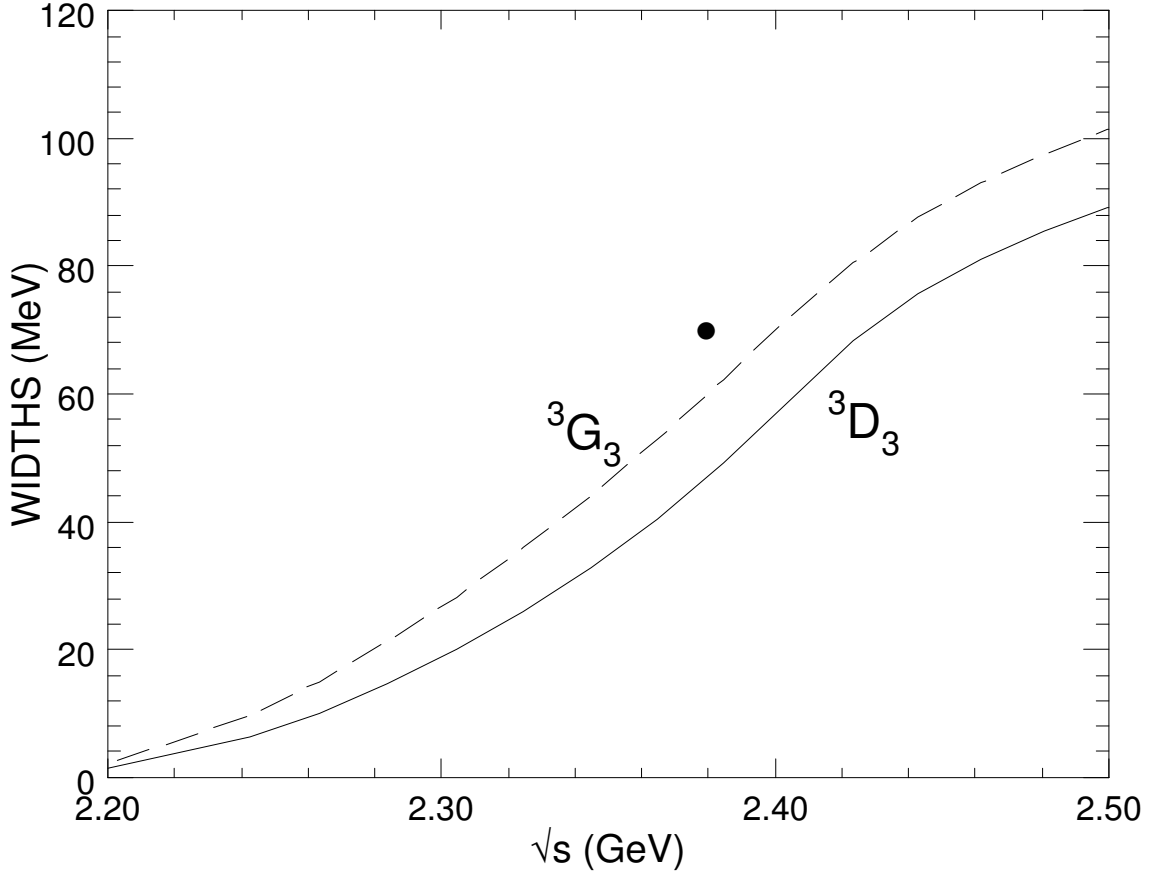}
\caption{The widths of the $^7S_3(\Delta\Delta)$ state in 
$I=0\;\; NN$ scattering: The solid curve initiates from $^3D_3(NN)$
and the dashed one from $^3G_3(NN)$. The bullet shows the energy
and width of the resonance reported e.g. in Ref. \cite{adlar90}.
\label{dgwid}
}
\end{figure}

Fig. \ref{dgwid} shows the widths for the most important
$^7S_3(\Delta\Delta)$ 
state coupled to the tensor-coupled $NN\;\; I=0$ system
$^3S_3- {^3G_3}$. It can be seen that at and below the 
two-$\Delta$ threshold the kinematic constraints with realistic
wave functions cause a drastic reduction in the width.
Actually at 2.38 GeV the more important $^3D_3$ wave would
get just about 50 MeV as the width, significantly less than 
the reported 70 MeV. Therefore, it seems that the narrowness
of the resonance cannot be used as an argument against the
possibility of its being of pure $\Delta\Delta$ origin.
The $^3G_3$ initiated state would have
13 MeV larger width, but its influence is suppressed by an 
order of magnitude due to the fact that to couple the $S$ and
$G$ waves one needs to operate twice by the tensor-like
transition potential. It may be possible to find dynamic
origins for further inelasticity, but as in the present 
phenomenological calculation its origin itself is not 
dynamically based, such a search would be inconsistent and
beyond the scope of the present work.

\begin{figure}[tb]
\includegraphics[width=\columnwidth]{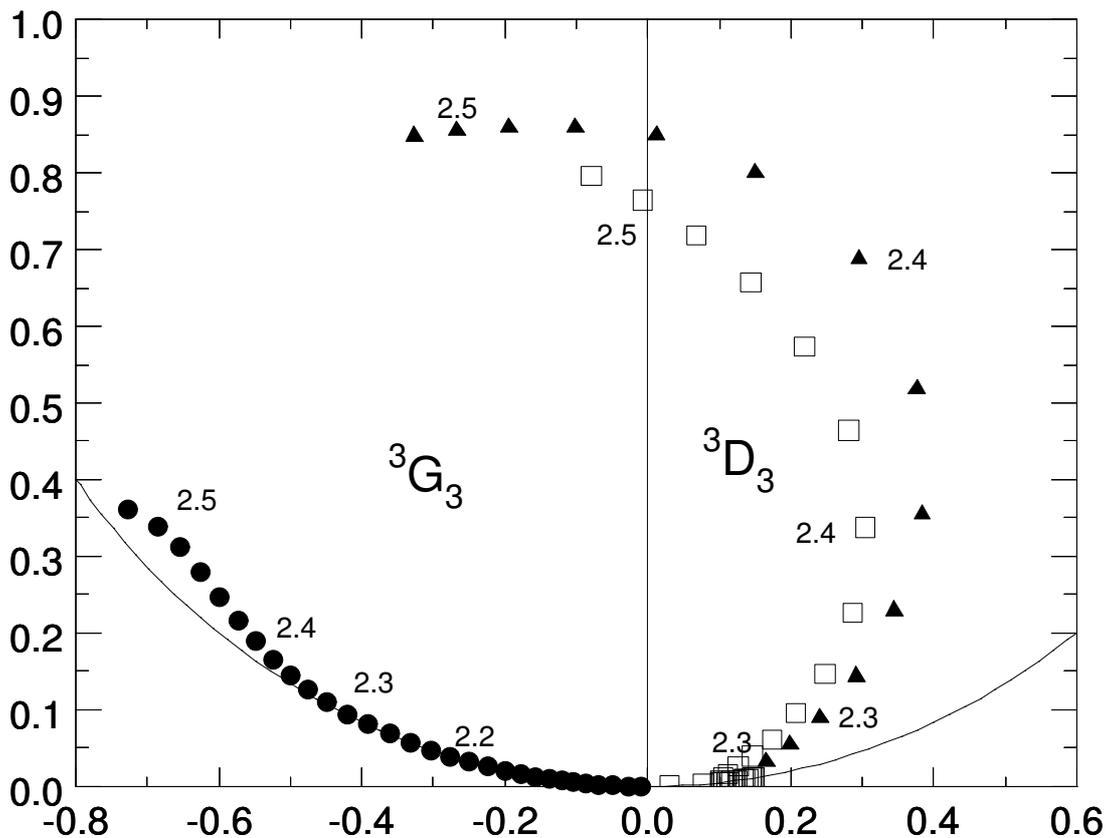}
\caption{Argand diagrams of two $NN$ partial waves with
a coupling to two $\Delta$'s. 
Squares $^3D_3$ and full circles $^3G_3$ states. The triangles
have three coupled $\Delta\Delta$ states and additional
attraction in each of these channels.
\label{argand3}
}
\end{figure}

Finally, Fig. \ref{argand3} shows the phases corresponding to
the initial $NN$ partial waves  $^3D_3$ and $^3G_3$ as Argand
diagrams. The open boxes are the results of a calculation
involving only the coupling to the $S$-wave $\Delta\Delta$.
For the $D$ wave the present diagram has curvature
indicative of a resonance but is significantly
more open than the result from the analysis of Ref. \cite{adlar90}
and remains mainly on the right-hand side up to the c.m. energy
of $\approx 2.5$ GeV. Understandably the threshold cusp should
appear rather at the double $\Delta$ mass 2.46 GeV in
agreement with the graph. (Actually the pole position of the
$\Delta$, 20 MeV lower, was used in the Schr\"odinger equation
to have the pole in the equivalent Lippmann-Schwinger
equation in its place.) Like in \cite{adlar90} there 
is a small nook on the unitarity circle
peaking at about 2 GeV followed by an ``armpit'' at 2.2 GeV.
This feature appears also as more pronounced in \cite{adlar90}.
In this calculation the $^3D_3$ phase shift remains remarkably
constant varying only smoothly between 3 to 5 degrees in the
energy range from 150 MeV to 1000 MeV (lab.). In Ref. \cite{arndt}
the phase should change sign above 800 MeV, but this change
does not appear in the later analyses \cite{arndt2,workman}, 
and in agreement with the latter result the phase shift remains 
well as constant up to 1 GeV. 
The resonance and threshold regions are still fairly
far above. Only well above 1 GeV in the laboratory energy an enhancement (together with inelasticity)
takes place. The phase shift maximizes at about 2.43 GeV (c.m.)
consistent with the doubled pole position.
The $G$-wave result is monotonous and
quite featureless and contrary to
Ref. \cite{adlar90} does not show any knot at 2.35 GeV.
Its phase grows nearly linearly with energy and the inelasticity
is small.

The weak tendency for a resonance behavior below 2.5 GeV
is somewhat puzzling considering that the width input in the 
coupled channels is only 50 MeV in the $^7S(\Delta\Delta)$ 
channel (at 2.38 GeV) and restricting presently to only this
single channel should rather favor a resonant behavior. As 
additional channels should bring more attraction, the next step
might be to include also the $^3D_3(\Delta\Delta)$ and 
$^7D_3(\Delta\Delta)$ components ($G$ waves should by far be 
negligible). A consistent calculation (adjusting also the
necessary extra repulsion in the $NN$ sector\footnote{The repulsion
in Eq. (\ref{modific}) is changed to $2700 \, e^{-6x}/x$,
practically only a range change.}) gives actually
slight smooth repulsion compared with the earlier one, 
negligible below $E_{\rm lab}\approx
1000$ MeV and 1--2 degrees in the resonance and threshold 
region. The overall effect is to smooth the threshold peaking,
since the effective threshold of these $\Delta\Delta\;D$
waves is significantly higher as discussed earlier. These
changes may be due to the fact that the width of the
$^7S_3(\Delta\Delta)$ state increases by about 10 MeV in this
calculation. In practice, the inclusion of the higher lying states does not change the position of the phase maximum at 2.43 GeV
appreciably.

One obvious and interesting possibility is an attractive
$\Delta\Delta$ interaction, which might bring the 
effective threshold
down to the $d'(2380)$ region. For this possibility a strong
artificial
test potential of about four pion strengths (in the $S$-wave
$NN$ potentials) is added in the $\Delta\Delta$ channels.
The effect is a faster and higher rise of the phase and
a subsequent faster fall after the phase maximum
 and a change of the sign already at
2.44 GeV. Also the position of the phase shift maximum is 
lowered close to 2.41 GeV. This result is shown in Fig. 
\ref{argand3} by triangles. Also the inelasticity is 
increased by this attraction though the widths themselves are 
not changed appreciably by this addition.

Adding such an extremely strong attraction is rather a drastic
act and one should question how such attraction could arise.
One might speculate about a crossed  two-pion exchange (with
the $\Delta$'s transforming to nucleons and back) being
attractive in high pion momentum parts. Each $N\Delta\pi$
vertex has about two times the $NN\pi$ coupling strength, 
so the strength from the
coupling coefficients alone could give a factor of 16 over the
normal $NN$ two-pion exchange (without $\Delta$'s). However,
comparisons with a real potential used here and expectations
based on that are not straightforward,
since unavoidably one meets on-shell pions with subsequent
imaginary parts. An actually dynamic calculation of the 
two-$\Delta$ width and an associated complex potential on the
same basis would be interesting.

\section{Conclusion}

The main conclusion of the present work is that the width of
the $\Delta(1232)$ resonance in a two baryon system $N\Delta$
or $\Delta\Delta$
is severely decreased due to the relative kinetic energy
of the baryons and their relative angular momentum. Since
the wave function is necessarily spatially confined, the 
expectation value of the kinetic energy is finite and out of use
for (internal) decay of the particles. Further, due to this 
wave function confinement the energy associated to the angular 
momentum barrier is quantized to finite average values, also to 
be subtracted from the energy available to internal excitations
and decays. Some obvious rules for the dependencies could be
seen in Fig. \ref{widths}. Firstly, even the largest of the
possible state dependent widths are significantly smaller than 
the free $\Delta$ width at the energy in question (corresponding 
to immobile baryons). The lowest angular momentum state
has the largest width. This is associated with effective
quantization of the above angular momentum energy, already
phenomenologically discussed for $I=1$ dibaryons in
Ref. \cite{dibarmass}. The higher orbital angular momentum 
$N\Delta$ or $\Delta\Delta$ states have increasingly smaller
widths. In $NN$ scattering with these intermediate states
also an important factor is whether $L_{N\Delta}$ or 
$L_{\Delta\Delta}$ is smaller or larger than the initial
$L_{NN}$ (or equal). In the first 
case the centrifugal barrier difference may partly cancel the
mass barrier $M_\Delta -M_N$ or $2(M_\Delta -M_N)$ thus
favoring the formation of the intermediate state,
and also in this case the width is larger than in cases where
the same intermediate state can be obtained from a lower $L_{NN}$.

As  seen in Fig. \ref{dgwid} the effective $\Delta\Delta$
width is significantly smaller than the single $\Delta$ width
at the relevant energies, at 2.38 GeV about 50 MeV, lower than
the reported $d'(2380)$ width 70 MeV. This result was
obtained with the most important $^7S_3(\Delta\Delta)$ alone. 
Including the $D$-wave $\Delta\Delta$'s increases 
the width to 60 MeV. Although this is just an input to 
an $NN$ scattering calculation, apparently 
an argument using just the $d'(2380)$ width vs. the 
free $\Delta$ width (not to say twice this)
is not necessarily assuring for its exotic origin. 

The use of the nonrelativistic Schr\"odinger equation might be
questioned in this calculation. 
Relativistic kinematics has been used to get
the center-of-mass $NN$ momentum and energy to meet
correctly the $N\Delta$ or $\Delta\Delta$ threshold. The subsequent 
nonrelativistic continuation should not, however, falsify
the above rather general and obvious results, which are not 
sensitive to this treatmentat least and in particular for the 
widths.

By this input alone one cannot obtain a resonant 
$^3D_3$ structure as low as 2.38 GeV, only at the 
$\Delta\Delta$ threshold (the calculated phase shift maximum 
at 2.43 GeV using the $\Delta$ pole position
as the mass). Adding arbitrarily as a test a strong attraction
of pion range it was possible to move the structure 
at least down to 2.41 GeV.
However, the question would remain about the origin of such strong
attraction, whether it could be hadronic (e.g. meson exchanges)
or possibly due to coupling to a genuine six-quark configuration.
Theoretically at least the $\Delta\Delta$ threshold should be 
there. Can one see two separate structures or have they merged
together as suggested by Bugg \cite{bugg1,bugg2} that 
resonances tend to synchronize together with thresholds?

\begin{acknowledgments}
I thank J. Haidenbauer, Ch. Hanhart, V. Komarov, T. L\"ahde,
A. Nogga and H. Machner for useful discussions 
and communications. I also 
acknowledge the kind hospitality of Forschungszentrum J\"ulich,
where much of this work was done.
\end{acknowledgments}

\end{document}